\documentclass[aps,amsfonts,nofootinbib,superscriptaddress,preprint]{revtex4-1}
\usepackage[normalem]{ulem} 
\makeatletter
\makeatletter
\usepackage{tikz}
\usetikzlibrary{decorations.pathmorphing, shapes.misc, shapes}
\usepackage{amsmath}
\usepackage{mdframed} 
\usepackage{tikz-feynman}
\tikzfeynmanset{compat=1.1.0}
\usepackage{tikz-feynman} 
\tikzfeynmanset{compat=1.1.0} 
\usepackage{feynmp-auto}
\usepackage{cleveref}
\usepackage{amssymb}
\usepackage{xcolor} 
\usepackage{tcolorbox} 
\usepackage{amsbsy}
\usepackage{bm}
\usepackage{epsfig} 
\usepackage{color}
\usepackage{slashed}
\usepackage{soul}
\usepackage{comment}
\usepackage{auncial}
\makeatletter
\usepackage{appendix}
\usepackage{epsfig} 
\usepackage{ulem}
\usepackage{pgfplots}
\newcommand{\stkout}[1]{\ifmmode\text{\sout{\ensuremath{#1}}}\else\sout{#1}\fi}
\newcommand{\ee}{\end{equation}}
\newcommand{\bb}{\begin{equation}}
\newcommand{\eqb}{\begin{eqnarray}}

\newcommand{\eqf}{\end{eqnarray}}

\usepackage[T1]{fontenc}
\usepackage{comment}
\usepackage[normalem]{ulem} 
\makeatletter
\usepackage{xcolor}
\usepackage{mdframed}
\usepackage{amsmath}
\usepackage{amssymb}
\usepackage{appendix}
\usepackage{amsbsy}
\usepackage{tikz}
\usepackage{booktabs}
\usetikzlibrary{decorations.pathmorphing, shapes.misc, shapes}
\usepackage{amsmath}
\usepackage{tikz}
\usepackage{feynmp-auto}
\usepackage{cleveref}
\usepackage{bm}
\usepackage{epsfig} 
\usepackage{color}
\usepackage{slashed}
\usepackage{soul}
\usepackage{comment}
\usepackage[normalem]{ulem} 
\makeatletter
\usepackage{amsmath}
\usepackage{amssymb}
\usepackage{appendix}
\usepackage{amsbsy}
\usepackage{bm}
\usepackage{epsfig} 
\usepackage{color}
\usepackage{slashed}
\usepackage{soul}
\usetikzlibrary{positioning,arrows,patterns,automata}
\usetikzlibrary{decorations}
\usetikzlibrary{trees}
\usetikzlibrary{decorations.pathmorphing}
\usetikzlibrary{decorations.markings}
\usetikzlibrary{calc}

\usepackage{feynmp-auto} 
\begin{document}
\title{The Role of Berry Phases in the QCD Vacuum Structure }
\author{J. Gamboa}
\email{jorge.gamboa@usach.cl}
\affiliation{Departamento de F\'{\i}sica, Universidad de Santiago de Chile, Santiago 9170020, Chile
}
\begin{abstract}
We revisit the origin of the vacuum angle $\theta$ in QCD using the adiabatic approximation combined with Fujikawa's method. By implementing a local chiral transformation and selecting a constant parameter $\alpha(x) = \theta$, we show that the QCD $\theta$-term emerges naturally in the effective action. This construction provides a non-perturbative interpretation of the axial anomaly and highlights the role of the adiabatic gauge condition in isolating the topological sector of the theory. As a consequence, the physical states acquire topological information encoded in the functional Berry phase $\Delta \alpha$, which manifests itself as a holonomy over the space of gauge configurations. This result offers a geometric and dynamical perspective on the structure of the QCD vacuum and its relation to anomaly-induced phases.
\end{abstract}

\maketitle

The vacuum angle $\theta$ is a prediction of QCD, derived from the topological structure of the space of gauge field configurations \cite{thooft1,review}. The term $\theta\, \text{Tr}(F_{\mu\nu} \tilde F^{\mu\nu})$ violates the CP symmetry and, among other effects \cite{jackiw}, induces an electric dipole moment for the neutron. The experimental non-observation of such a CP violation implies that $\theta \ll 1$, which in turn means that the neutron electric dipole moment must be extremely small, though not necessarily zero. Therefore, $\theta$ remains a physical prediction of QCD, whose detection depends solely the experimental sensitivity \cite{dipole}.

The most widely accepted solution to this problem is the Peccei–Quinn proposal \cite{PQ1}, which explains the smallness of the violation of CP in QCD by promoting the parameter $\theta$ to a dynamical pseudo-scalar field, denoted by \(a(x)\), called the \emph{axion} \cite{weinberg,wilczek,sikivie}. This field has a small mass and couples to the gauge field via the interaction
\[
\frac{a(x)}{f_a}\, \text{Tr}(F_{\mu\nu} \tilde F^{\mu\nu}),
\]
where $f_a$ is a coupling constant with dimensions of energy. The effective dynamics of the axion is governed by the Lagrangian
\[
\mathcal{L} = \frac{1}{2} (\partial_\mu a)^2 - \frac{1}{2} m_a^2 a^2 + \frac{a(x)}{f_a}\, \text{Tr}(F_{\mu\nu} \tilde F^{\mu\nu}) + \cdots,
\]
where $m_a$ is the axion mass, generated by non-perturbative effects. In this framework, the vacuum expectation value of the field $a$ adjusts dynamically to cancel the $\theta$-term, thus resolving the strong CP problem \cite{review12}.

However, one may approach the problem from a different perspective and consider that QCD is a gauge theory in which chiral anomalies play a central role \cite{bell,adler}. Instead of starting from the standard generating functional, one may write the functional directly by incorporating the effect of the fermionic Jacobian, following Fujikawa's method \cite{fujikawa}
\begin{equation}
Z = \int {\cal D} A_\mu\, {\cal D} \bar{\psi}\, {\cal D} \psi\, J(\alpha)\, e^{-S}, \label{jab1}
\end{equation}
where $J(\alpha)$ is Fujikawa’s Jacobian, which captures the non-invariance of the fermionic measure under a local chiral transformation, and has the explicit form
\begin{equation}
J(\alpha) = \exp\left( - \frac{g^2}{16 \pi^2} \int d^4 x\, \alpha(x)\, \text{Tr}(F_{\mu \nu} \tilde F^{\mu \nu}) \right).
\end{equation}
The transformed Euclidean action then reads:
\begin{equation}
S = \int d^4x \left( \frac{1}{4} F_{\mu\nu} F_{\mu\nu} + \bar{\psi} \left( \slashed{D}[A] + i \gamma_5\, \slashed{\partial} \alpha(x) \right)\psi - \frac{g^2}{16 \pi^2} \alpha(x)\, \text{Tr}(F_{\mu \nu} \tilde F^{\mu \nu})\right).
\end{equation}

We now propose a conceptual departure from the standard treatment of CP violation in QCD. We note that:
\begin{enumerate}
\item The chiral anomaly is a non-perturbative effect that manifests at all scales, regardless of the energy regime considered.

\item Suppose that we are interested in the low-energy regime of QCD and apply the adiabatic approximation.
\end{enumerate}

This latter point implies that we can modify the fermionic kinetic term via the replacement \cite{gamboa}
\[
\slashed{A} \to \slashed{A} + \slashed{\cal A},
\]
where ${\cal A}_\mu$ is a non-Abelian Berry connection \cite{berry,wz} associated with the slow evolution of the gauge background. As a consequence, the effective action takes the form:
\begin{equation}
S = \int d^4x \left( \frac{1}{4} \text{Tr}(F_{\mu\nu} F^{\mu\nu}) + \bar{\psi} \left( \slashed{D}[A] + \slashed{\cal A} + i \gamma_5\, \slashed{\partial} \alpha(x) \right)\psi - \frac{g^2}{16 \pi^2} \alpha(x)\, \text{Tr}(F_{\mu \nu} \tilde F^{\mu \nu})\right). \label{accion1}
\end{equation}

This equation tells us that $\frac{\delta S}{\delta \alpha} = 0$ implies 
$\langle \partial \cdot J_5 \rangle \sim \text{Tr}(\tilde F F)$; that is, the independence of the gauge-fixed action with respect to $\alpha$ encodes the chiral anomaly.

On the other hand, if we choose the \emph{adiabatic gauge}, defined by the condition \cite{gamboa}
\begin{equation}
\partial_\mu \alpha = i \gamma_5 {\cal A}_\mu, \label{adua1}
\end{equation}
then the action (\ref{accion1}) becomes:
\begin{equation}
S = \int d^4x \left( \frac{1}{4} \text{Tr}(F_{\mu\nu} F^{\mu\nu}) + \bar{\psi} \left( \slashed{D}[A]  \right)\psi - \frac{g^2}{16 \pi^2} \alpha(x)\, \text{Tr}(F_{\mu \nu} \tilde F^{\mu \nu})\right). \label{accion12}
\end{equation}
This directly yields the QCD vacuum angle if $\alpha(x) = \theta = \text{const.}$; that is, the QCD vacuum arises as a consequence of a global chiral transformation.

However, one must still analyze the implications of the gauge condition (\ref{adua1}), which upon integration gives the total accumulated phase:
\begin{equation}
\Delta \alpha = i \int dx^\mu\, \gamma_5 {\cal A}_\mu. \label{acum1}
\end{equation}
Although in the adiabatic gauge the limit $\alpha(x) \to \theta = \text{const}$ eliminates the explicit derivative term $\partial_\mu \alpha$, and allows the QCD topological term to be derived from a global chiral transformation, this procedure comes at a price: the functional Berry phase accumulated in (\ref{adua1}) modifies the structure of the physical state space. This is an expected effect that reflects the topologically nontrivial nature of the QCD vacuum \cite{thooft1}.

This phase--which arises as the holonomy of a connection over the functional space of gauge configurations-- does not directly influence the appearance of the $\theta$-term, but it does determine how the quantum states of the system must be defined. In particular, in the presence of this Berry connection, physical states cannot be regarded simply as plane waves; rather, they must be constructed as sections parallel transported by $\mathcal{A}_\mu$, that is, as states modified by the corresponding holonomy matrix.

This structure reflects the fact that the Hilbert space acquires a nontrivial fibration, with ``dressed'' states that incorporate topological information about the functional trajectory followed \cite{fk}. Therefore, the functional Berry phase acts as a bridge between the global topological structure of the field configuration space and the local definition of the physically observable states.

The author thanks Fernando Méndez for his valuable and insightful discussions over the course of this work. This research was supported by DICYT (USACH), grant number 042531GR$\_$REG.

\end{document}